\documentclass[aps,prl,superscriptaddress,amsfonts,amsmath,amssymb,twocolumn,showpacs,floatfix,footinbib]{revtex4-1}

\usepackage{psfrag,graphicx}
\usepackage{dcolumn}
\usepackage{amsmath,amssymb}
\usepackage{bm}
\usepackage{amsfonts,amssymb,amsmath}        
\usepackage{multirow}

\usepackage{epstopdf}

\newcommand{\be}{\begin{equation}}
\newcommand{\ee}{\end{equation}}
\newcommand{\bq}{\begin{eqnarray}}
\newcommand{\eq}{\end{eqnarray}}

\newcommand{\ket}[1]{\left | \, #1 \right\rangle}	

\newcommand{\rf}[1]{(\ref{#1})}

\allowdisplaybreaks[4]

\begin{document}

\title{Realizing all $so(N)_1$ quantum criticalities in symmetry protected cluster models}

\author{Ville Lahtinen}	
\affiliation{Dahlem Center for Complex Quantum Systems, Freie Universitat Berlin, 14195 Berlin, Germany}		
\affiliation{Institute for Theoretical Physics, University of Amsterdam, Science Park 904, 1090 GL Amsterdam, The Netherlands}

\author{Eddy Ardonne}			
\affiliation{Department of Physics, Stockholm University, AlbaNova University Center, SE-106 91 Stockholm, Sweden}

\date{\today}

\begin{abstract}

We show that all $so(N)_1$ universality class quantum criticalities emerge when one-dimensional generalized cluster models are perturbed with Ising or Zeeman terms. Each critical point is described by a low-energy theory of $N$ linearly dispersing fermions, whose spectrum we show to precisely match the prediction by $so(N)_1$ conformal field theory. Furthermore, by an explicit construction we show that all the cluster models are dual to non-locally coupled transverse field Ising chains, with the universality of the $so(N)_1$ criticality manifesting itself as $N$ of these chains becoming critical. This duality also reveals that the symmetry protection of cluster models arises from the underlying Ising symmetries and it enables the identification of local representations for the primary fields of the $so(N)_1$ conformal field theories. For the simplest and experimentally most realistic case that corresponds to the original one-dimensional cluster model with local three-spin interactions, our results show that the $su(2)_2 \simeq so(3)_1$ Wess-Zumino-Witten model can emerge in a local, translationally invariant and Jordan-Wigner solvable spin-1/2 model. 

\end{abstract}


\maketitle

A striking property of quantum phase transitions is that of universality. Near a critical regime separating distinct quantum states of matter, the system specific microscopic details are lost and the system acquires universal behavior that is characterized by scaling exponents of distinct observables near the transition \cite{book:sachdev}. A crucial step in putting order to the zoo of distinct universalities was made by noting that the diverging correlation length at the critical point implies conformal invariance, and thus a description by a conformal field theory (CFT) \cite{bpz84,byb}. This approach is particularly powerful in one spatial dimension (1D), where the relevant CFT fully describes (the scaling of) the correlation functions.

1D quantum models with critical points are thus the natural playgrounds for quantum criticality. The simplest and the most celebrated ones are the transverse field Ising (TFI) chain \cite{epw70}, that at criticality is described by the Ising CFT, and the XY chain \cite{lsm61} where the anisotropic transition is described by the so called $so(2)_1$ CFT. So ubiquitous are these universality classes that criticality beyond them is often filed under ``exotic''. Still, considering this ubiquitous nature these simple models have played in various fields of physics, it was only rather recently when the details of the criticality of the TFI chain were experimentally probed \cite{Coldea10}. The challenge of probing quantum criticality beyond these simple models owes to the lack of tractable models that admit accessible experimental realization. To go beyond them, often higher spins \cite{Tu08,Alet11,Tu11,Orus11,Capponi13,Tu13}, infinite range couplings \cite{haldane88,shastry88}, broken translational invariance \cite{Lahtinen14}, strongly interacting systems \cite{zamolodchkov80,Huijse15,Rahmani15}, or combinations there of \cite{michaud13,nielsen11}, are required. Here we show that a hierarchy of local, exactly solvable and translationally invariant spin-1/2 models provides the simplest setting to generalize the Ising universality class by realizing all $so(N)_1$ criticalities, that were previously considered in manifestly $so(N)$ symmetric higher spin chains \cite{Tu08,Tu11} or in models with broken translational invariance \cite{Lahtinen14}. Furthermore, we show that the universality manifests itself microscopically through every quantum critical point being dual to $N$ non-locally coupled critical transverse field Ising chains.  

Our hierarchy can be viewed as a generalization of 1D cluster models \cite{Doherty09} -- a class of stabilizer models with a symmetry protected degenerate ground state manifold \cite{Son11,Smacchia11} -- whose ground states in 2D were first proposed as a universal resource for one-way quantum computation \cite{Briegel01}. While in 1D these models are universal resources only for a single qubit, they have proven accessible settings to study the entanglement \cite{Pachos04,Doherty09,Skrovseth09,Son11,Smacchia11,Montes12,Cui13,Giampaolo14,Bridgeman15} and the computational power \cite{Else12,Miller15,Miller15-1} of symmetry protected states, as well as the robustness of edge states in a many-body localized phase \cite{Bahri15}. Perturbing the pure stabilizer models with Ising and Zeeman terms, we define the generalized cluster models with periodic boundary conditions by the Hamiltonians (the original 1D cluster model corresponds to $A=3$)
\be \label{HN}
  H^{(A)}_{\rm cluster}  =  \sum_{i=0}^{L-1}  \left( \right. C^A_i + J \sigma^x_{i} \sigma^x_{i+1} + h \sigma^z_i \left. \right),
\ee
where $C^A_i = \sigma^y_{i} \sigma^z_{i+1} \cdots \sigma^z_{i+(A-2)} \sigma^y_{i+(A-1)}$ acts on $A$ adjacent sites and $J$ and $h$ are the magnitudes of the Ising and Zeeman terms, respectively. The cluster stabilisers $C^A_i$ commute with each other and thus for $J=h=0$ one obtains the cluster state as the unique ground state that satisfies $C^A_i \ket{\Psi}=-\ket{\Psi}$ for all $i$. Anologous to the spin-1 Haldane phase \cite{Affleck87,Nijs89,Pollmann12}, these states are symmetry protected topological (SPT) phases protected by global $Z_2^{\times A-1}$ symmetries $\mathcal{Z}_n=\prod_{i=0}^{(L-1)/(A-1)} \sigma^z_{(A-1)i+n}$ for $n=0,\ldots,A-2$, and characterized by a string order parameter and edge states \cite{Son11,Smacchia11,Bahri14}. We show below that even if these symmetries are broken, the perturbed Hamiltonians can still exhibit global $Z_2^{\times N}$ symmetries, where $N$ depends on the perturbation in question.

\begin{figure}[t]
\begin{tabular}{cc}
\includegraphics[width=0.60\columnwidth]{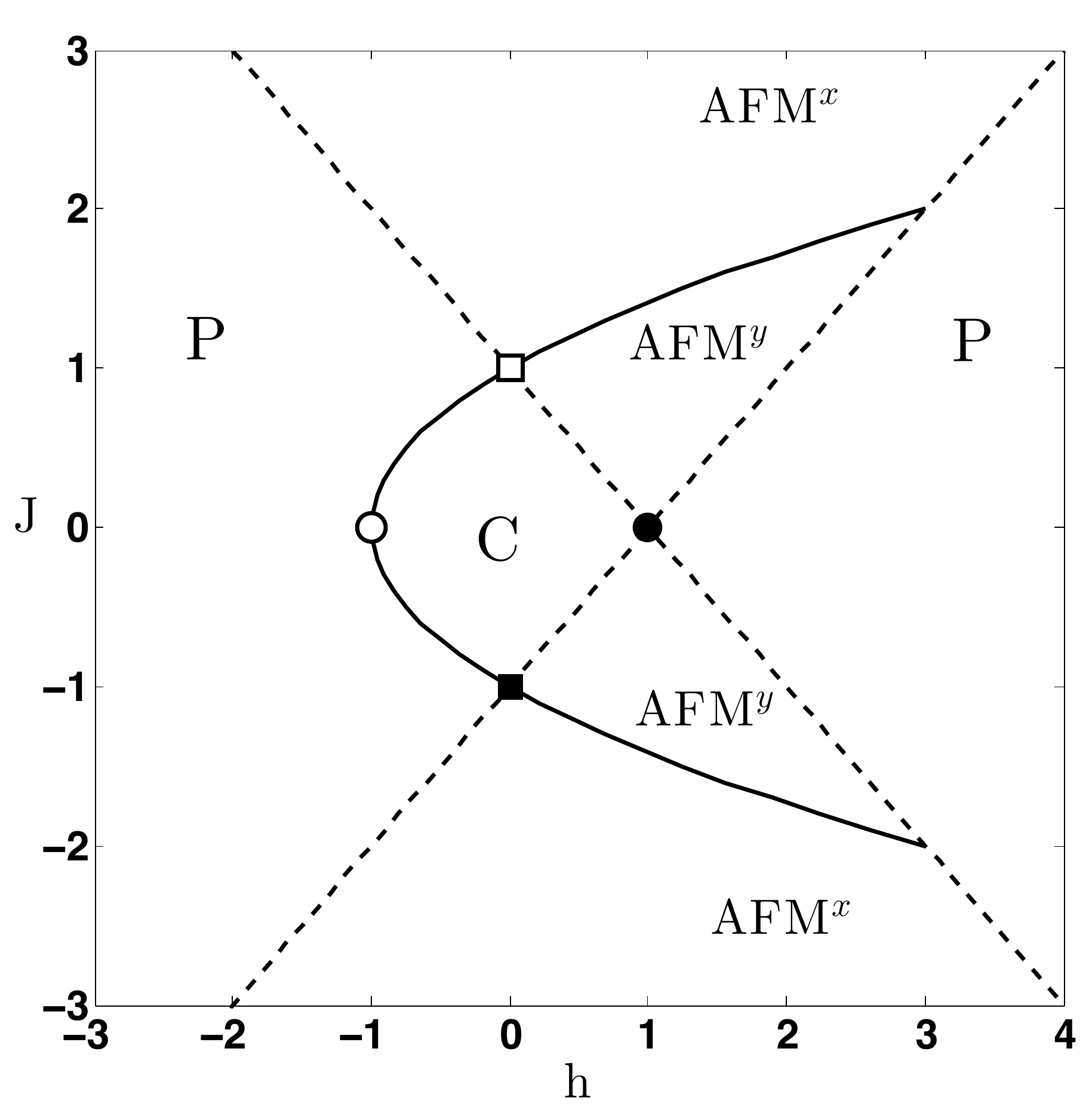} & \includegraphics[width=0.38\columnwidth]{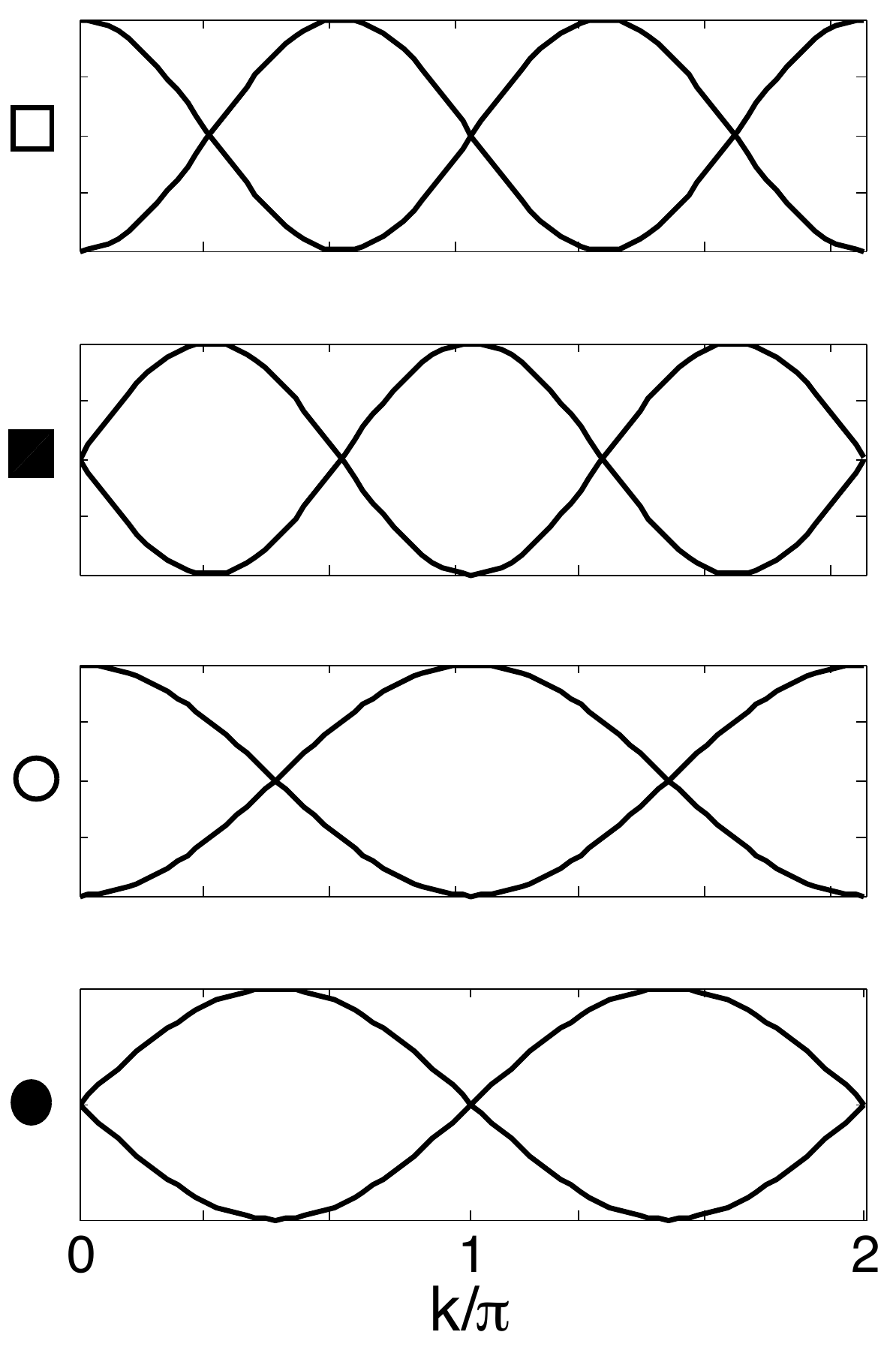}
\end{tabular}
\caption{Phase diagram of the $3$-cluster model exhibiting gapped cluster ($C$), polarised ($P$) and anti-ferromagnetic ($AFM^{x,y}$) phases \cite{Montes12}. The solid (dashed) lines indicate second order quantum phase transitions along which the spectral gap closes at two (one) distinct Fermi momenta. The insets show the spectrum \rf{E} at the multi-critical points $(J,h)=(\pm1,0)$ and $(J,h)=(0,\pm1)$ where the critical spectrum exhibits two (circles) and three (squares) Majorana cones, respectively. }
\label{PD3}
\end{figure}

To study the phase diagrams of the $A$-cluster models \rf{HN}, we map \rf{HN} to a problem of free fermions by introducing two Majorana fermions per physical site
\be \label{Maj_mapping}
	a_{i} = \left( \prod_{k < i} \sigma^z_k \right) \sigma^x_i, \qquad b_{i} = \left( \prod_{k < i} \sigma^z_k \right) \sigma^y_i.
\ee
Assuming a chain of $L$ sites with periodic boundary conditions, in momentum space the Hamiltonian takes the form $H^{(A)}_{\rm cluster}
	=i \sum_{k} \left[  \epsilon_k a_k^\dagger b_k - \epsilon_k^* b_{k}^\dagger a_{k} \right]$, where $\epsilon_k = -h + J \exp(-ik) + \exp[i(A-1)k]$ and the Majorana operators satisfy in the momentum space $c_k^\dagger=c_{-k}$. The momentum takes the values $k=(2m+1-P)\pi/L$, where $m=0,\ldots,L-1$ and $P=\pm 1$ is the eigenvalue of the parity operator $\mathcal{P}_{cluster}=\prod_{i=0}^{L-1} \sigma^z_i$. The spectrum can be brought to the diagonal form $H=\sum_k |\epsilon_k| (2 \gamma_{k}^\dagger \gamma_{k} -1)$ by moving to the complex fermion basis $\gamma_{k} = \frac{1}{\sqrt{2}} \left(  i \frac{|\epsilon_k|}{\epsilon^*_k} a_k + b_{k} \right)$
with the eigenvalues given by
\bq \label{E}
  |\epsilon_k| & = & \left[ 1+h^2+J^2 + 2J \cos(Ak) \right. \nonumber \\
  \ & \ & \ \ \left. -2h(J \cos(k)+ \cos((A-1)k)) \right]^{1/2}.
\eq
The resulting phase diagram for the simplest 3-cluster model, that was first studied in Ref \cite{Montes12}, is shown in Figure \ref{PD3}. While the phase diagrams become in general more complex for $A>3$ (presented in the Supplementary Material \cite{supp}), they all share a common feature: One always finds multi-critical points at $(J,h)=(\pm 1,0)$ and $(J,h)=(0,\pm 1)$, where there is a second order quantum phase transition from the cluster phase to an anti-ferromagnetic or a spin polarised phase, respectively. Determining the CFT describing these critical points for general $A$ is our main result.

To gain insight into these critical points, we focus first on $(J,h)=(\pm 1,0)$ where the dispersion reduces to $|\epsilon_k| = \sqrt{2 \pm 2\cos(A k)}$. This vanishes at the $N=A$ Fermi points $K_n=\pi/N+2\pi n/N$ ($J=1$) or $K_n=2\pi n/N$ ($J=-1$), where $n=0,\ldots,N-1$. Expanding around each Fermi momenta by writing $k=K_n+p$ with $p \ll 1$, one finds linear dispersion $|\epsilon_{K_n+p}|=Np + \mathcal{O}(p^2)$ implying a low-energy description in terms of $N$ fermions of velocity $N$. Carrying out a similar analysis for the $(J,h)=(0,\pm1)$ critical points, one finds $N = A-1$ Fermi points with linearly vanishing dispersion, as illustrated in Figure \ref{PD3}. 
It is well known that at low energy, the spectrum of the critical TFI chain is described by a single linearly dispersing fermion and that the criticality is described by the Ising CFT with central charge $c=1/2$ \cite{byb}. Thus the low-energy description in terms of $N$ fermions with a linear dispersion naively suggests that the corresponding critical points are described by a product theory of $N$ Ising CFTs with $c=N/2$. Indeed, central charges of $c=1$ and $c=3/2$ have been obtained for the 3-cluster model at $(J,h)=(0,\pm 1)$ \cite{Bridgeman15} and $(J,h)=(\pm 1,0)$ \cite{Smacchia11}, respectively. 

However, neither the number of linearly dispersing fermions nor the central charge uniquely fix the CFT. In addition to the product CFT Ising$^{\times N}$, the central charge of $c=N/2$ and the spectrum of $N$ fermions are also consistent with the so called $so(N)_1$ CFTs \cite{Tu08,Alet11,Tu11,Lahtinen14}. As these theories have dramatically different primary field content \cite{byb,supp}, they can be distinguished by the energy levels and degeneracies in the finite-size energy spectrum that is fully determined by the field content of the corresponding CFT. For non-chiral models such as ours, the spectrum of an $L$ site chain takes the form \cite{byb}
\be \label{CFTpred}
	E = E_0 L - \frac{\pi v c}{6L} + \frac{2\pi v}{L}(2h_{\alpha} +n),
\ee
where the on-site energy $E_0$ and the velocity $v$ are non-universal numbers and $n$ is a non-negative integer. On the other hand, the central charge $c$ and the scaling dimension $h_{\alpha}$ of each primary field $\alpha$ are universal and determined by the CFT. The spectrum can be simplified by setting the ground state energy to $E=0$ and scaling the spectrum such that the first excited state has energy $E=2h_{\alpha'}$, with $h_{\alpha'}$ being the smallest non-zero scaling dimension. The spectrum assumes then the simple form $E=2h_{\alpha}+n$ where the energies arrange themselves into integers offset by the scaling dimensions that characterise the CFT. The predictions of the energy levels and degeneracies by both $so(N)_1$ and Ising$^{\times N}$ CFTs are presented in the Supplementary Material \cite{supp}. In Figure \ref{ECFT3}, we plot the spectra for the 3-cluster model at the critical points $(J,h)=(1,0)$ and $(0,1)$, which match precisely the prediction by the $so(3)_1$ and $so(2)_1$ CFTs, respectively. We have verified that this structure holds for general $A$-cluster models, which leads to our main result: The multi-critical points $(J,h)=(\pm 1,0)$ and $(\pm 1,0)$ of every $A$-cluster model have low-energy theories in terms $N=A$ and $N=A-1$ fermions, respectively, and they are described by $so(N)_1$ CFTs. 

\begin{figure}[t]
\includegraphics[width=\columnwidth]{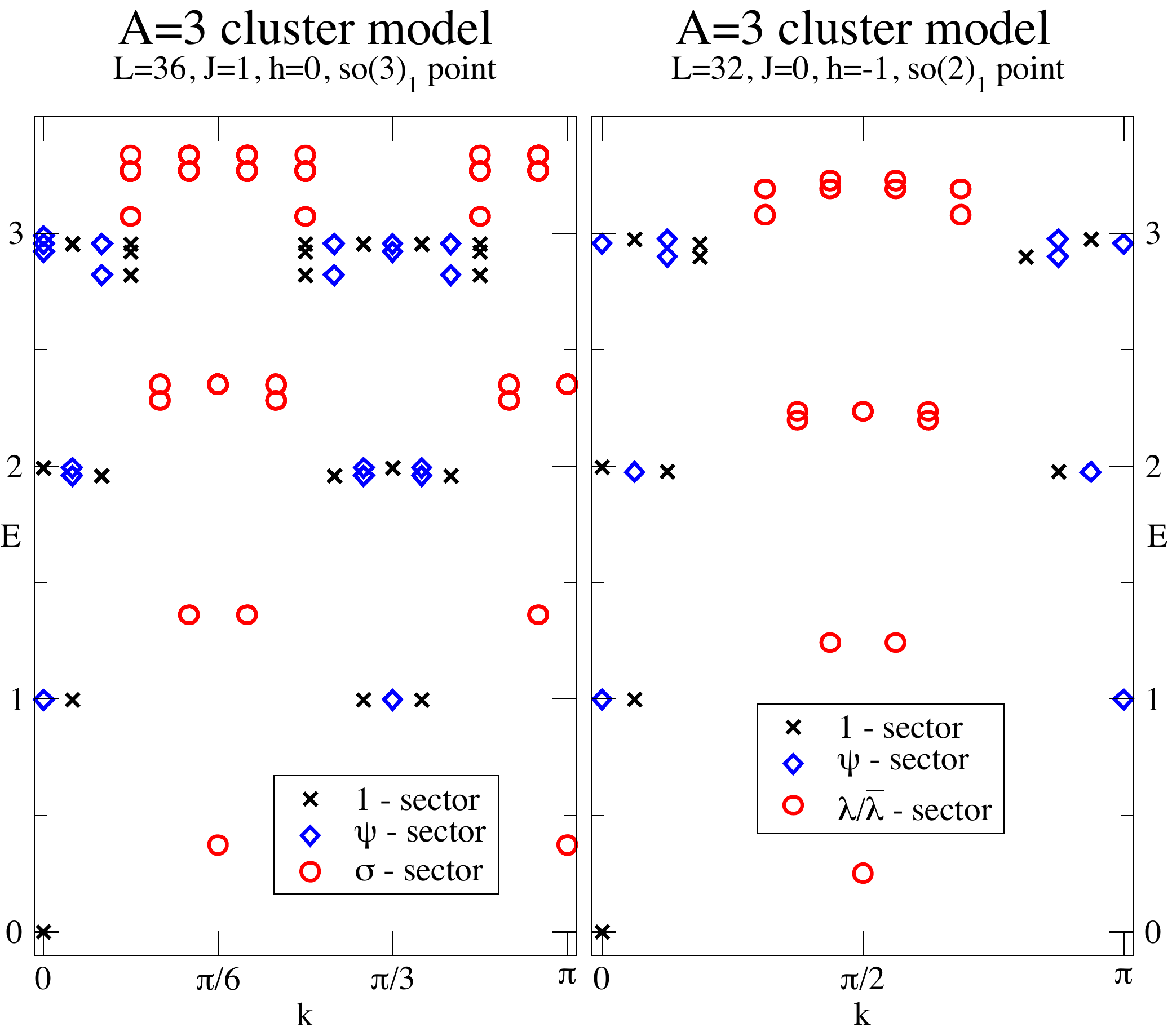}
\caption{Matching of the CFT spectra $E=2h_\alpha+n$ for the $so(3)_1$ and $so(2)_1$ critical points of the $N=3$ cluster model. The $so(3)_1$ CFT has three primary fields $\{1,\psi,\sigma \}$ with scaling dimensions $h_1=0$, $h_\psi=1/2$ and $h_\sigma=3/16$, while the $so(2)_1$ CFT has four primary fields $\{1,\psi,\lambda,\bar{\lambda} \}$ with scaling dimensions $h_1=0$, $h_\psi=1/2$ and $h_\lambda=h_{\bar{\lambda}}=1/8$. In both cases the rescaled energies follow the predicted pattern \rf{CFTpred}, with the degeneracies (not indicated) at each energy level matching those predicted for $so(N)_1$ CFTs, see \cite{Lahtinen14}. }
\label{ECFT3}
\end{figure}

To analytically demonstrate the universality of these critical points, we now turn to show that at each $so(N)_1$ critical point the system can be mapped to $N$ {\it non-locally coupled} TFI chains with $Z_2^{\times N}$ symmetry. These symmetries correspond to different decompositions of the parity sectors $\mathcal{P}_{cluster}=\prod_{n=1}^{A-1} \mathcal{Z}_n=\prod_{n=1}^N \mathcal{P}_n$ into hidden symmetry sectors labeled by $\mathcal{P}_n = \pm 1$. The canonical spin duality transformations we employ are inspired by Refs \cite{Lahtinen14,Doherty09} and exist for general $N$ (presented in the Supplementary Material \cite{supp}). For the sake of clarity we focus here on the $3$-cluster model. When $h=0$ we employ first the canonical transformations
\begin{align} \label{dualJ}
\sigma^{y}_{3j} & =  \mathcal{Q}^<_j \tau^{x}_{3j+1}, & \sigma^{z}_{3j} & =  \tau^{y}_{3j} \tau^{y}_{3j+1} \mathcal{Q}^>_j, \\ 
\sigma^{y}_{3j+1} & =  -\tau^{x}_{3j} \tau^{x}_{3j+1} \tau^{x}_{3j+2}, & \sigma^{z}_{3j+1} & = \mathcal{Q}^<_j \tau^{z}_{3j} \tau^{x}_{3j+1} \tau^{x}_{3j+2} \mathcal{Q}^>_j \nonumber \\
\sigma^{y}_{3j+2} & =  \tau^{x}_{3j+2} \mathcal{Q}^>_j, & \sigma^{z}_{3j+2} & =  \mathcal{Q}^<_j \tau^{y}_{3j} \tau^{y}_{3j+2}, \nonumber 
\end{align}
where $\mathcal{Q}^<_j=\prod_{i<j} (\tau^{z}_{3i} \tau^{z}_{3i+2})$ and $\mathcal{Q}^>_j=\prod_{i>j} (\tau^{z}_{3i} \tau^{z}_{3i+1})$. Assuming that $L$ is a multiple of three, applied to \rf{HN} we obtain
\begin{align} 
	H^{(3)}_{\rm cluster} = \sum_{j=0}^{L/3-1} \big( & J \tau^x_{3j} \tau^x_{3j+3} & + \tau^z_{3j} & + J \mathcal{P}_1 \mathcal{P}_2 \tau^x_{L-3} \tau^x_{0} + \nonumber   \\
	\ & \tau^x_{3j+1} \tau^x_{3j+4} &  +J \tau^z_{3j+1} & + \mathcal{P}_0 \mathcal{P}_2 \tau^x_{L-2} \tau^x_{1} +  \nonumber \\
	 \ & \tau^x_{3j+2} \tau^x_{3j+5} & +J \tau^z_{3j+2} & + \mathcal{P}_0 \mathcal{P}_1 \tau^x_{L-1} \tau^x_{2}  \big), 
\end{align} 
where the duals of $\mathcal{P}_n=\prod_{m=0}^{L/3-1} \tau^z_{3m+n}$ are hidden global $Z_2^{\times 3}$ symmetries of the perturbed cluster model (all the $\mathcal{Z}_n$ symmetries are broken by $J \neq 0$). A simpler duality exists for $h \neq 0$, but $J=0$ when the $\mathcal{Z}_n$ symmetries are preserved. Assuming that $L$ is even, we introduce 
\be \label{dualh}
	\sigma^y_{2j} = \mathcal{Q}_j^< \tau^y_{2j}, \quad \sigma^y_{2j+1} = \mathcal{Q}_j^> \tau^y_{2j+1}, \quad \sigma^z_{i} = \tau^z_{i},
\ee
where now $\mathcal{Q}_j^<= \prod_{i<j} \tau^z_{2i+1}$ and $\mathcal{Q}_j^>= \prod_{i>j} \tau^z_{2i}$, which directly give
\begin{align}
	H^{(3)}_{\rm cluster} = \sum_{j=0}^{L/2-1} \big( &  \tau^y_{2j} \tau^y_{2j+2} & + h \tau^z_{2j} & + \mathcal{P}_1 \tau^y_{L-2} \tau^y_{0} + \nonumber   \\
	\ & \tau^y_{2j+1} \tau^y_{2j+3} & + h  \tau^z_{2j+1} & + \mathcal{P}_0 \tau^y_{L-2} \tau^y_{1}, 
\end{align} 
where $\mathcal{P}_n=\prod_{m=0}^{L/2-1} \tau^z_{2m+n}$ now coincide with the $\mathcal{Z}_n$ symmetries. 

Along both cuts of the phase diagram that contain the multi-critical points we thus find the same structure: The $A$-cluster models with Ising or Zeeman perturbations are dual to TFI chains that are non-locally coupled such that the boundary condition for the $n$th TFI chain depends on the product $\prod_{m \neq n}\mathcal{P}_m$ of the Ising symmetry sectors of all the other chains. This coupling is precisely of the general form that, based on the framework of condensate-induced transitions between topological phases \cite{bais09}, has been recently shown to change the criticality between Ising$^{\times N}$ and $so(N)_1$ universality classes \cite{Lahtinen14}. In the sector where $\mathcal{P}_n=1$ for all $n$, or for open chains with free boundary conditions, the spectrum is indeed equal to decoupled TFI chains and at criticality the spectrum is indistinguishable from that of Ising$^{\times N}$ CFT. The crucial point is that this sector is only a single subsector of the $\mathcal{P}_{cluster}=1$ symmetry sector. To construct the full $so(N)_1$ critical spectrum, one needs to appropriately combine the spectra of $N$ critical TFI chains using all Ising symmetry sectors. This subtle mixing of the spectra from different symmetry sectors underlies the emergence of the $so(N)_1$ instead of the Ising$^{\times N}$ universality class and highlights the need to consider all symmetry sectors when identifying the relevant CFT by spectral means.

\begin{figure}[t]
\includegraphics[width=\columnwidth]{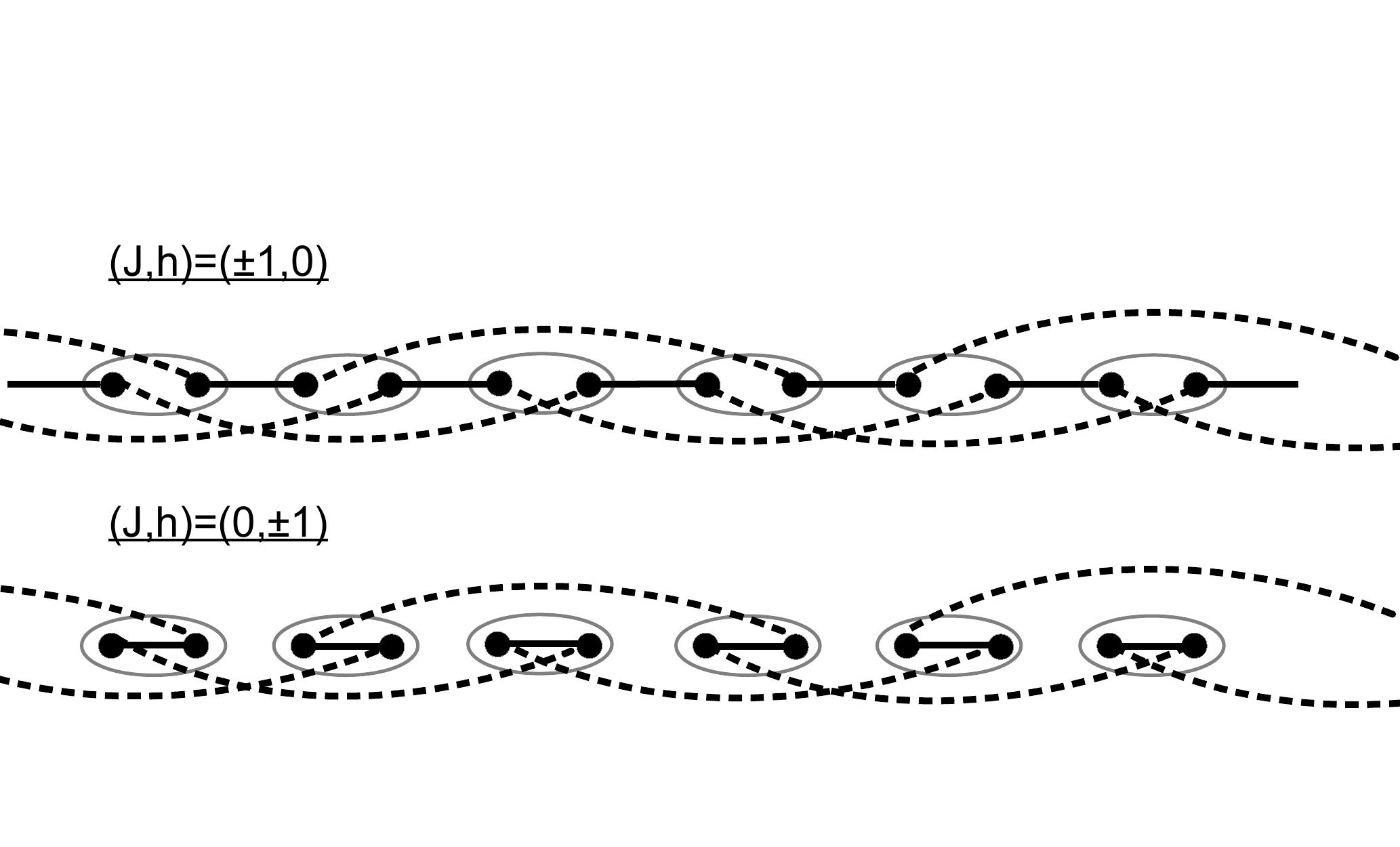}
\caption{The Majorana representation of the $H^{(3)}_{cluster}$ cluster model. The ellipses denote the physical spin sites that are represented by two Majoranas. The solid lines correspond to Majorana terms originating from $\sigma_i^x \sigma^x_{i+1}=i b_i a_{i+1}$ or $\sigma^z_i=-i a_i b_i$ terms, while dashed lines correspond to cluster terms $\sigma_i^y \sigma^z_{i+1} \sigma^y_{i+2}=i a_i b_{i+2}$. Along the $h=0$ ($J=0$) cut the system decouples locally into three (two) Majorana wires. The wires are coupled non-locally though through their boundary conditions that for each chain are either periodic or anti-periodic depending on $\mathcal{P}_{cluster}= \prod_j (i a_j b_j)=\pm 1$.}
\label{Majchains}
\end{figure}

These results can be generalized to the critical points between arbitrary cluster phases. Consider the critical Hamiltonian $H=\sum_i ( C_{\alpha,i}^A + C_{\beta,i}^B)$, where we have introduced two non-commuting cluster stabilizers $C_{\alpha,i}^A = \sigma^\alpha_{i} \sigma^z_{i+1} \cdots \sigma^z_{i+(A-2)} \sigma^\alpha_{i+(A-1)}$ with $\alpha=x,y$ (we define $C^1_{\alpha,i}= -\sigma_i^z$ ). Employing the mapping to Majorana fermions \rf{Maj_mapping}, one immediately finds that for $\alpha \neq \beta$ the system maps into $|(A-1)+(B-1)|$ Majorana chains, which are only coupled through their boundary conditions given by $\mathcal{P}_{cluster}$  (see Figure \ref{Majchains} for the Majorana representation of the 3-cluster model with $A=3$ and $B=1$ or $2$). For $\alpha=\beta$ one finds $|A-B|$ such chains. Since Majorana chains are well known to be dual to TFI chains, their coupling through their boundary conditions translates precisely into the non-local coupling between TFI chains discussed above. Thus we conclude that the critical points between arbitrary cluster phases are always described by the $so(N)_1$ CFT with $N=|(A-1) \pm (B-1)|$. That all critical points due to arbitrary competing cluster stabilizers can be mapped to the Hamiltonian
\be \label{HsoN1}
	H_{so(N)_1}=\sum_{i=1}^{L-N} \tau^x_i \tau^x_{i+N} + \sum_{n=1}^N (\prod_{m \neq n} \mathcal{P}_m)\tau^x_{L-N+n} \tau^x_n + \sum_{i=1}^L \tau_i^z
\ee 
with $Z_2^{\times N}$ symmetry, is the microscopic manifestation of the universality of $so(N)_1$ criticality and provides a full classification of criticalities in this set of exactly solvable models.

This universal microscopic description can be employed to identify candidates for the spin representations for fields related to the chiral primaries $\psi$ and $\sigma$ of the $so(N)_1$ CFT with scaling dimensions $h_\psi=1/2$ and $h_{\sigma}=N/16$, respectively. If a field with scaling dimension $\Delta_\alpha = 2 h_\alpha$ is represented by a local operator $O_i^\alpha$, then the ground state correlator at the critical point is expected to decay as $\langle O^\alpha_i O_{j}^\alpha \rangle \sim |i-j|^{-2\Delta_\alpha}$. For a critical TFI chain, which can be viewed as the special $N=1$ case of the hierarchy, the non-chiral combinations of the $\psi$ ($h_\psi=1/2$) and $\sigma$ ($h_\sigma=1/16$) primaries are related to local operators through $\psi_L\psi_R \sim \tau^z$ and $\sigma_L\sigma_R \sim \tau^x$ (the chiral left (L) and right (R) moving parts require non-local operators)\cite{byb}. Thus local operators in the $A$-cluster models with the scaling $h=1$ of the $\psi_L\psi_R$ field are given by the duals of the $\tau^z_j$ operators, while a natural candidate for a local operator with scaling dimension $2N/16$ of $\sigma_L\sigma_R$ is the dual of the product of $N$ adjacent $\tau^x_j$ operators. For the $so(3)_1$ critical point of the 3-cluster model at $(J,h)=(\pm1,0)$ these would explicitly be given by, for instance, $\psi_L\psi_R \sim \sigma^x_j\sigma^x_{j+1}$ and  $\sigma_L\sigma_R \sim \sigma^y_j$. We leave the explicit verification, either analytically \cite{epw70,lsm61} or numerically \cite{Mong14-1,Stojevic14,Bridgeman15} for future work. 
 
Finally, we show that the dual picture in terms of $N$ boundary coupled TFI chains is consistent with and useful also to study the perturbed gapped cluster phases \cite{Doherty09,Son11,Smacchia11}. For $J=h=0$ and for periodic boundary conditions, for $A$ odd (even) there holds $\prod_{i=1}^L C_i^A=\prod_{i=1}^L \sigma_i^z$ ($1$), which implies $L$ ($L-1$) independent constraints and thus a unique (two-fold degenerate) ground state. In the cluster phase $A-1$ TFI chains are in the ferromagnetic phase, but the unique ground state for $A$ odd arises from the boundary coupling enforcing that there is only one sector ($\mathcal{P}_n=1$ for all $n$) where all chains have periodic boundary conditions (the lowest energy contribution per TFI chain), while for $A$ even this occurs for two sectors ($\mathcal{P}_n=\pm 1$ for all $n$). For free boundary conditions the boundary coupling is removed and there are only $L-(A-1)$ stabiliser constraints. This gives rise to a $2^{A-1}$ fold ground state degeneracy that is consistent with $A-1$ out of the $N$ unconstrained TFI chains being in the ferromagnetic phase. While the $\mathcal{Z}_n$ symmetries protecting the pure cluster state are in general broken by Zeeman, Ising or competing cluster term perturbations, the dual TFI chain picture with hidden $Z_2^{\times N}$ symmetry persists throughout the cluster phase. Since this picture enables to isolate the hidden degrees of freedom that contribute to the properties of perturbed cluster states (regardless of $N$ only $A-1$ TFI chains contribute to the $A$-cluster state), it can be valuable for studying the computational power of symmetry protected topological states \cite{Else12,Miller15,Miller15-1}.

We have shown that transitions between symmetry protected cluster phases realize all quantum critical points in the universality class of $so(N)_1$ CFTs. We explicitly demonstrated the microscopic universality of the transitions by mapping every multi-critical point to $N$ non-locally coupled TFI chains with $Z_2^{\times N}$ symmetry. Thus in addition to being of interest to quantum information \cite{Pachos04,Doherty09,Skrovseth09,Son11,Smacchia11,Montes12,Cui13,Giampaolo14,Bridgeman15,Else12,Miller15,Miller15-1}, cluster models are also accessible platforms for probing quantum criticality beyond Ising universality class that can emerge between distinct SPT states. Experimental realizations of these models has been proposed in optical lattices \cite{Pachos04} and with trapped ions \cite{Lanyon11}. Furthermore, an important corollary of our results is the discovery that the $su(2)_2 \simeq so(3)_1$ criticality, which has previously been discovered only in spin-1 or higher systems \cite{zamolodchkov80,Tu08,nielsen11,michaud13}, can emerge also in a translationally invariant spin-1/2 chain with local interactions. A full classification of criticalities between SPT phases beyond the $Z_2$ protection of the integrable cluster models remains an interesting open question.


{\it Acknowledgements}. -- V.L. acknowledges the support by the Dutch Science Foundation NWO/FOM and the Dahlem Research School POINT Fellowship program. E.A. acknowledges support from the Swedish Research Council.

\clearpage

\appendix
\section{Supplementary Material}

In the supplementary material we present the phase diagrams and give the duality transformations for general $A$-cluster models. We also briefly review how the CFT predicts the spectrum at a critical point and give the predictions by $so(N)_1$ and Ising$^{\times N}$ CFTs.

\section{Phase diagrams for arbitrary $A$-cluster models}

The spectrum of the $A$-cluster model is in general given by
\bq
  |\epsilon_k| & = & \left[ 1+h^2+J^2 + 2J \cos(Ak) \right. \nonumber \\
  \ & \ & \ \ \left. - 2h(J \cos(k)+ \cos((A-1)k)) \right]^{1/2}.
\eq
To find the parameters $J$ and $h$ for which the gap closes, we complete squares to obtain
$|\epsilon_k|^2 = \bigl( h - J \cos(k) - \cos((A-1)k) \bigr)^2 + 
\bigl(J \sin(k) - \sin((A-1)k)\bigr)^2$.
We find that there is gap closing point at $k=0$ when $J = h-1$ and at
$k=\pi$ when $J = -h - (-1)^{A}$. In addition, the gap closes along a curve in $(h,J)$, that can
be parametrized by the momentum $0\leq p \leq \pi$ for which $\epsilon_p = 0$. This curve is given by
\be
 (h,J) (p)= \bigl(\frac{\sin(A p)}{\sin(p)},\frac{\sin((A-1)p)}{\sin(p)} \bigr).
\ee 
Along this curve, the gap closes
at both $k=p$ and $k=2\pi-p$. The multi-critical points at $(h,J)=(\pm1,0)$ and $(h,J)=(0, \pm1)$ occur when this line intersects itself and/or the two lines with a gap closing at a fixed momentum. 

\begin{figure}[h]
\begin{tabular}{cc}
\includegraphics[width=4.2cm]{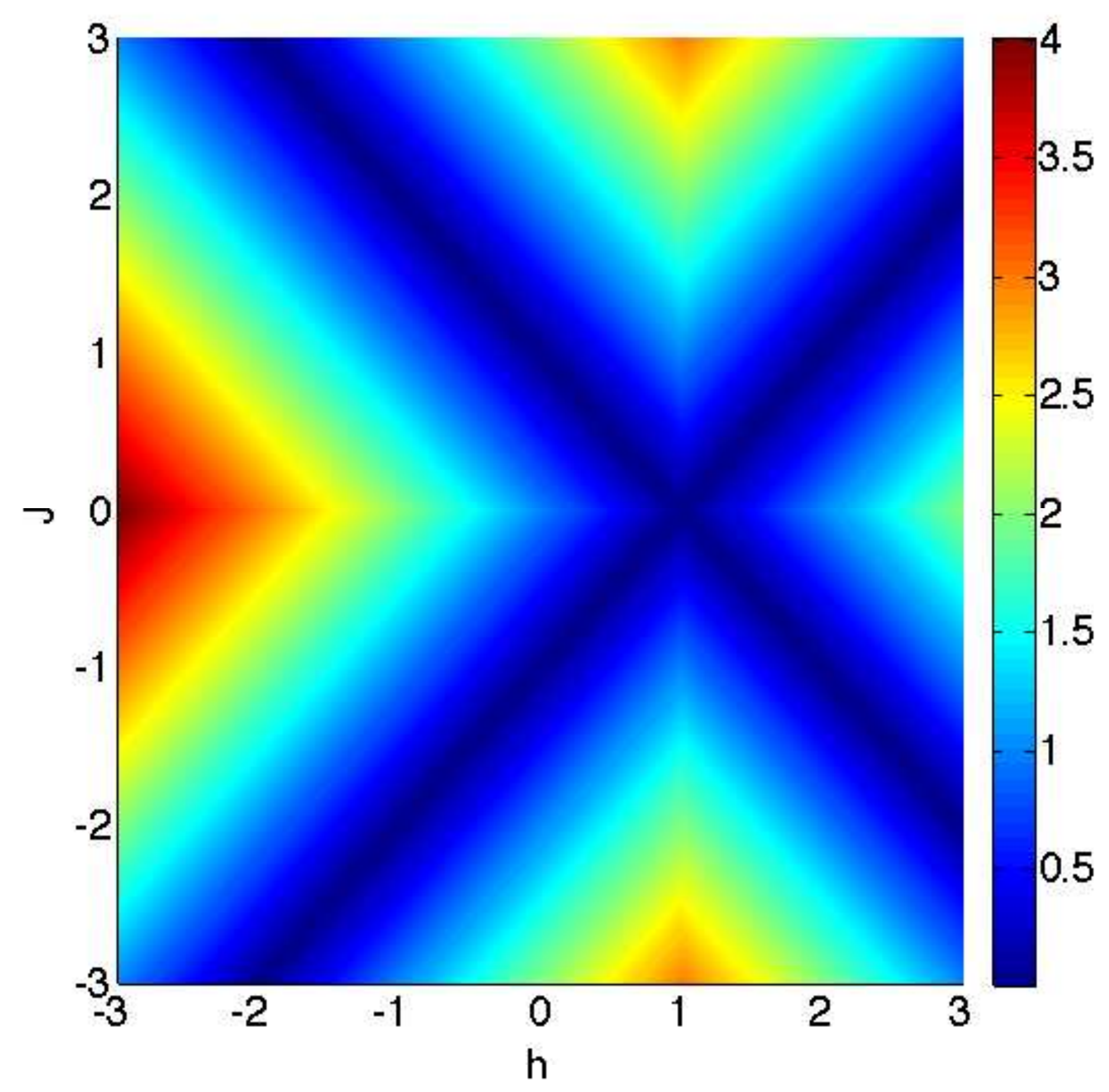} & \includegraphics[width=4.2cm]{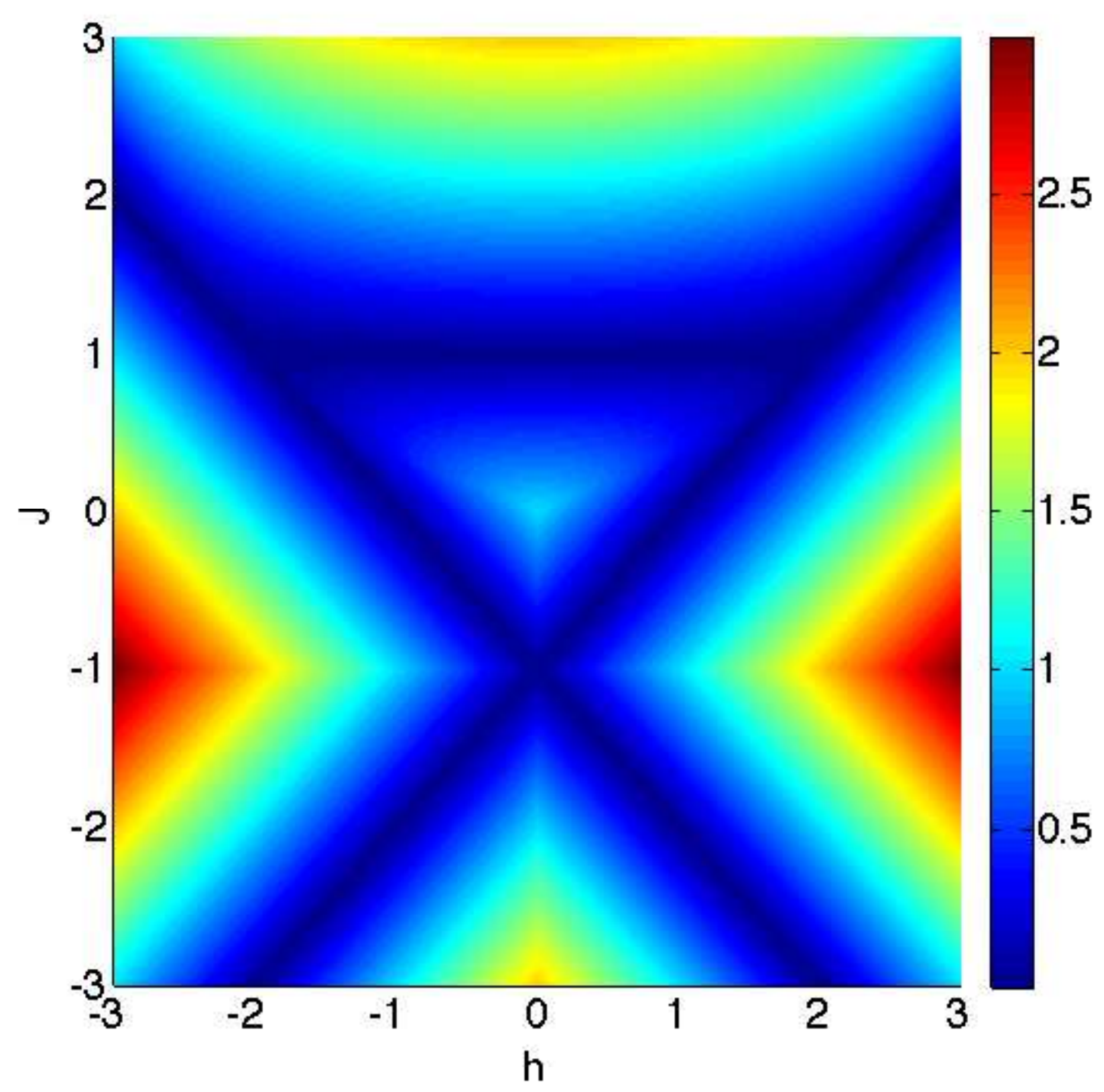} \\
$A=1$ & $A=2$ \\
\includegraphics[width=4.2cm]{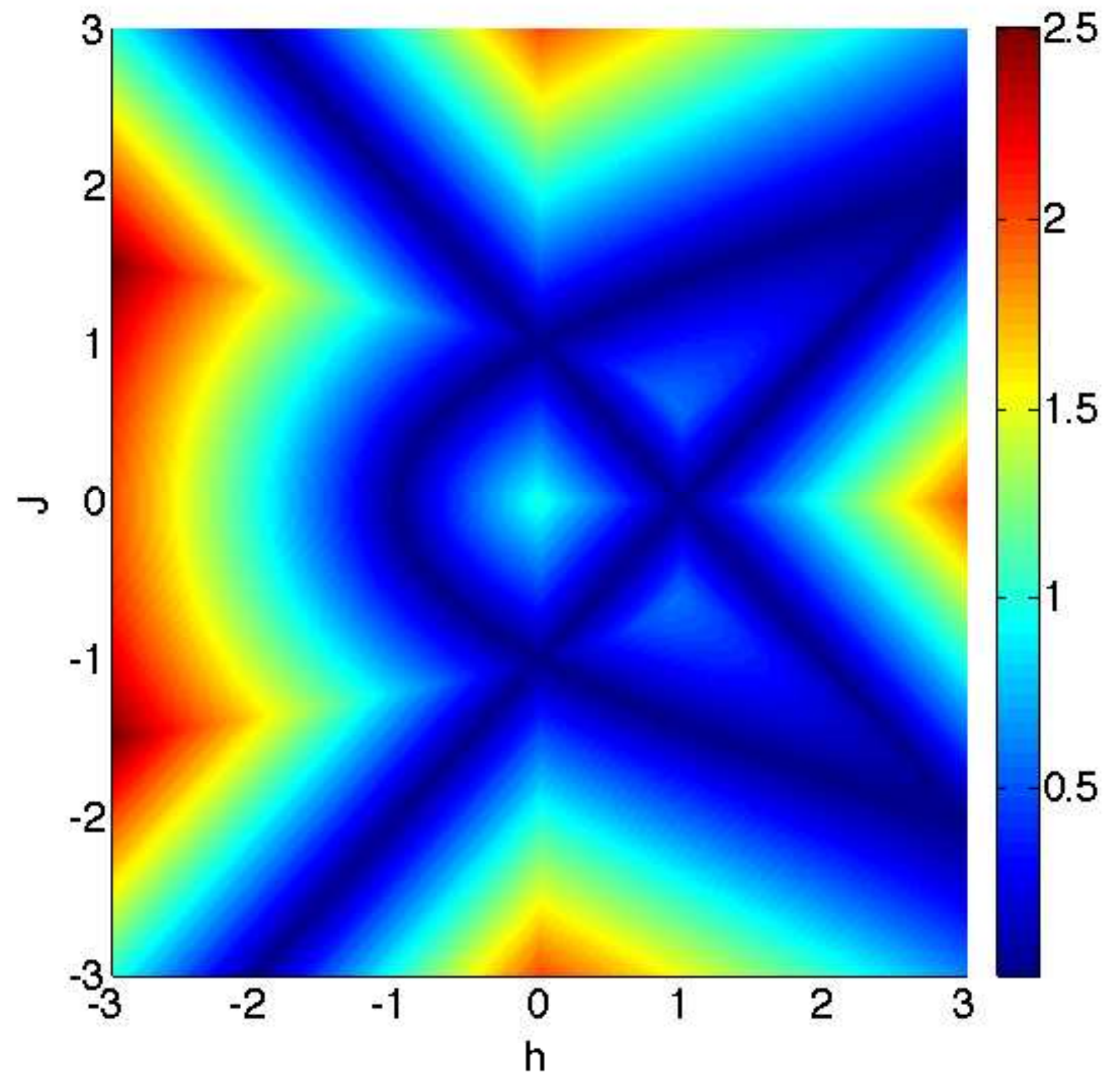} & \includegraphics[width=4.2cm]{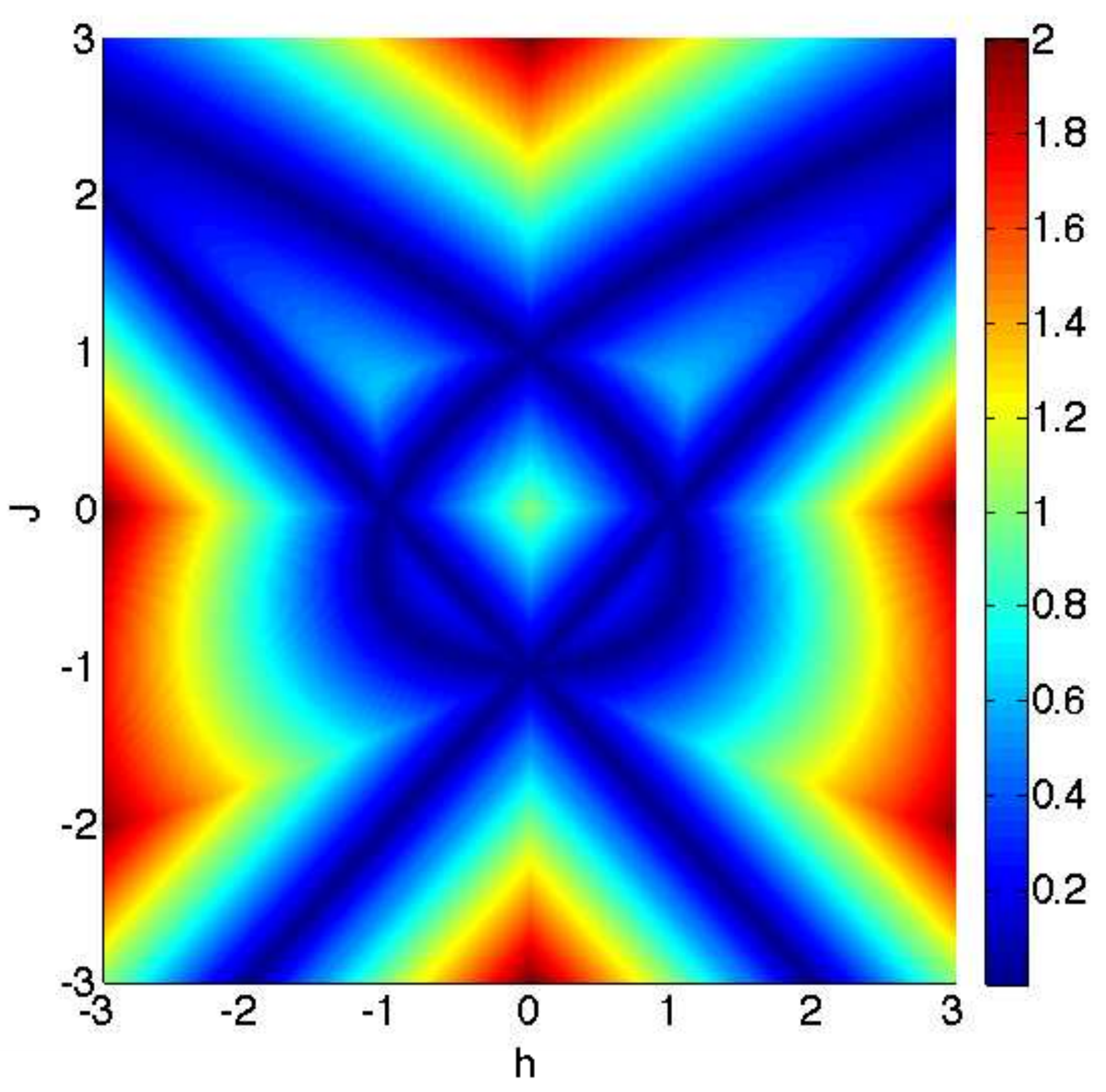} \\
$A=3$ & $A=4$ \\
\includegraphics[width=4.2cm]{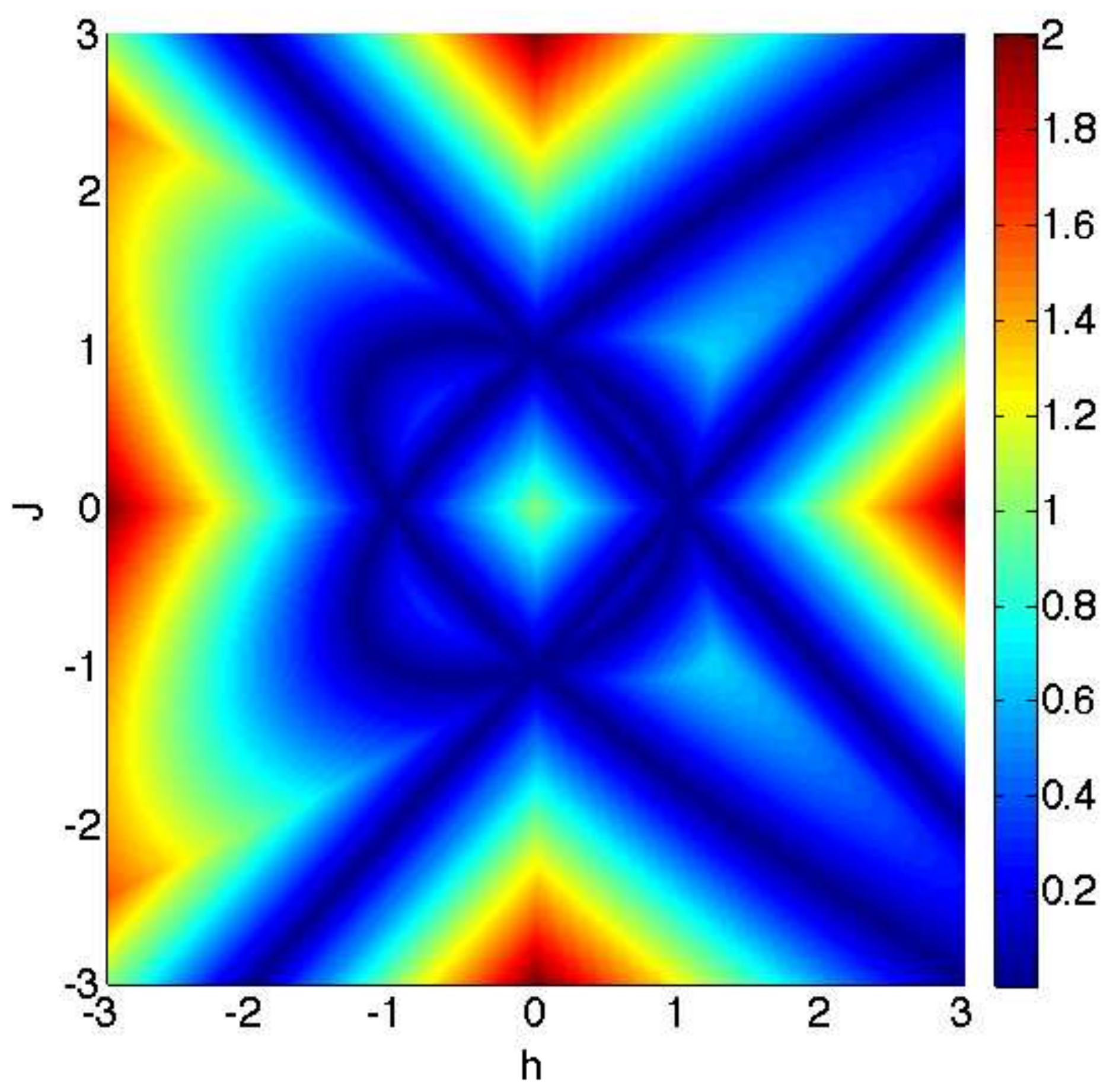} & \includegraphics[width=4.2cm]{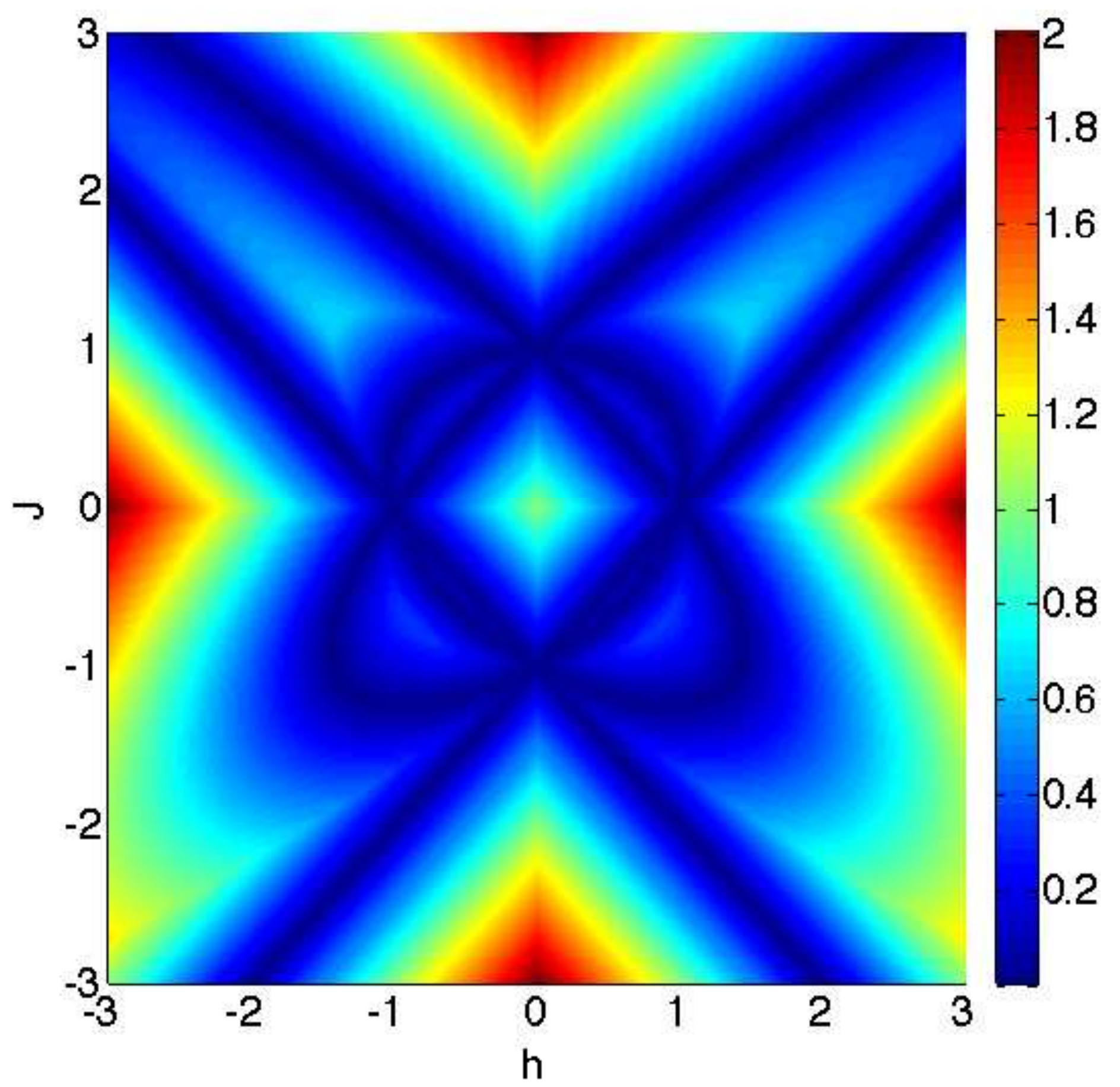} \\
$A=5$ & $A=6$ \\
\includegraphics[width=4.2cm]{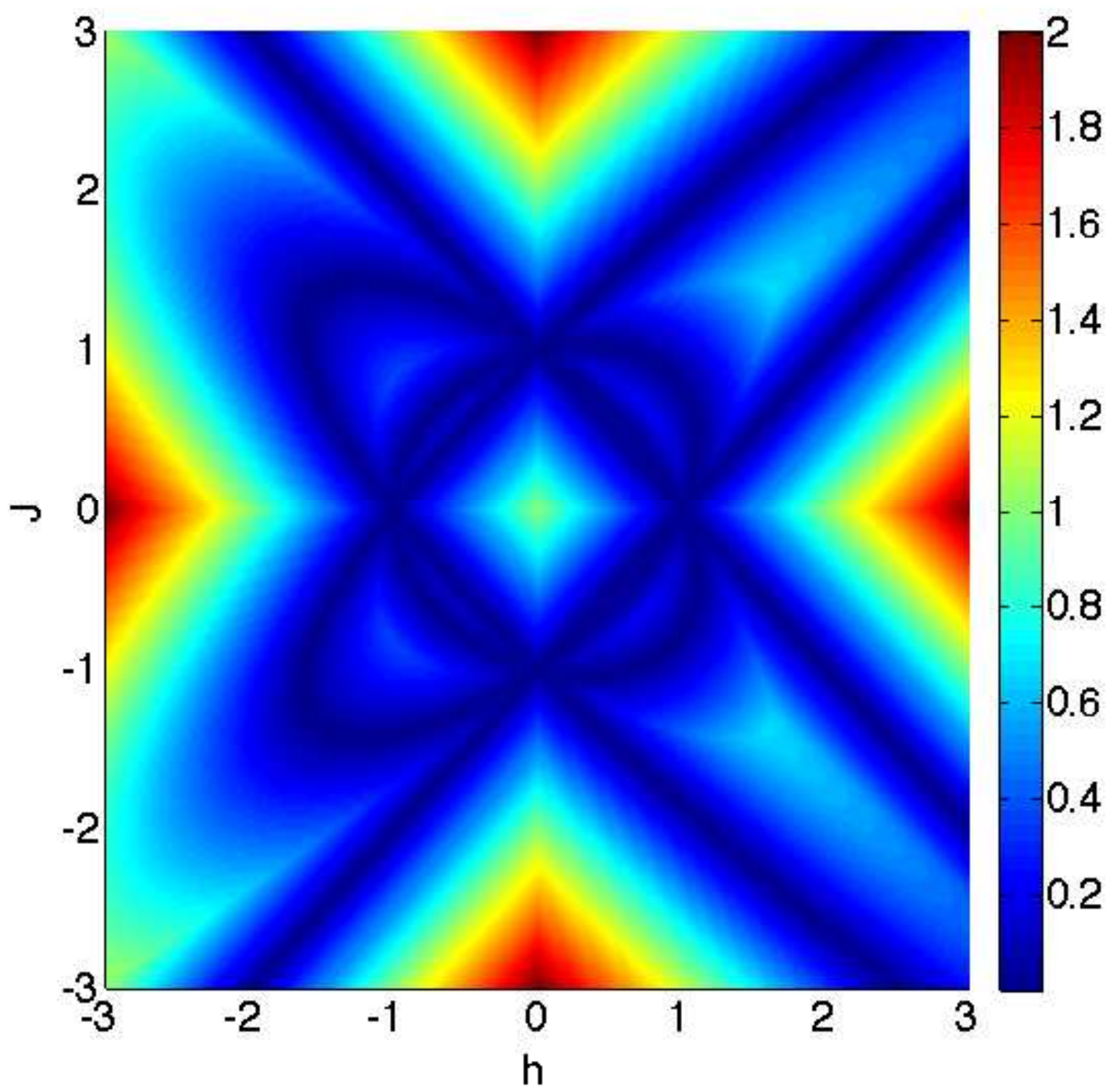} & \includegraphics[width=4.2cm]{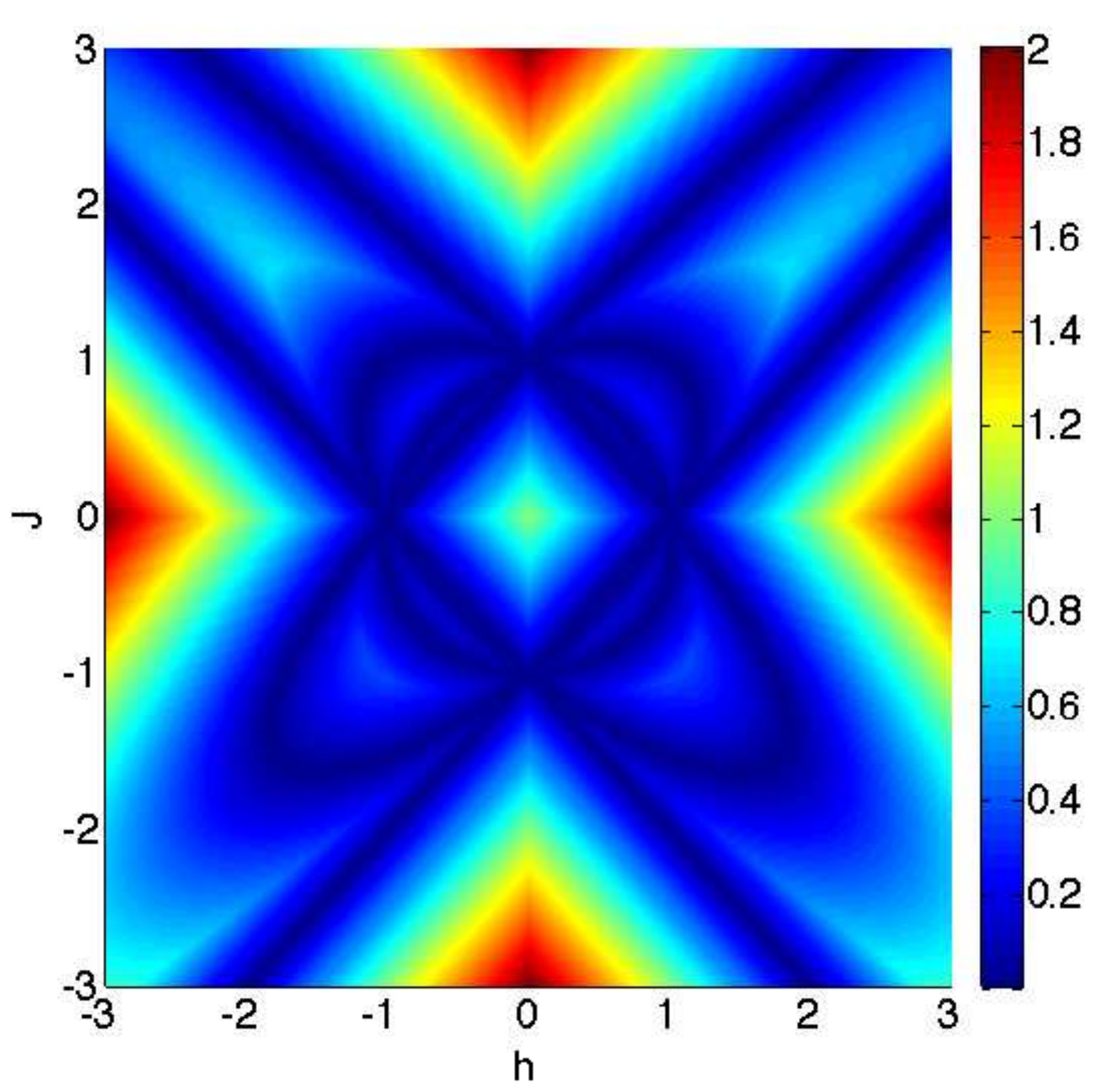} \\
$A=7$ & $A=8$ 
\end{tabular}
\caption{Phase diagrams for general $A$-cluster models with colour encoding the magnitude of the spectral gap. The cases $A=1$ and $A=2$ correspond to the transverse field Ising and XY chains, respectively. For $A>2$ The cluster phase occurs always around $J,h \ll 1$, while the (anti-)ferromagnetic and the polarized phases occur again for $J \gg h,1$ and $h \gg J,1$, respectively. The fine-tuned incommensurate phases arising from the competition between all three terms are gapped magnetic phases. }
\label{PDall}
\end{figure}

The phase diagrams for some of the smallest $A$ are illustrated in Figure \ref{PDall}. As $A$ increases, the $A$-cluster phase becomes sharply bounded by $J+h<1$. Several fine-tuned gapped phases also emerge due to the competition between all the three Hamiltonian terms. These are magnetic phases similar to the incommensurate anti-ferromagnetic phases of the $3$-cluster model studied in Ref \cite{Montes12}.

\section{Duality transformations for arbitrary $A$-cluster models}

A system of $N$ locally decoupled TFI chains can be realised on a single 1D system by considering a single TFI chain with $N$th nearest neighbour interactions only. When they are coupled together non-locally such that the boundary condition (periodic or anti-periodic) of chain $n$ depends on the product $\prod_{m \neq n} \mathcal{P}_m$ of the Ising symmetry sectors of all other chains, it has been argued in Ref \cite{Lahtinen14} that the criticality, when all the TFI chains are simultaneously critical, changes from the Ising$^{\times N}$ to the $so(N)_1$ universality class. With the aid of duality transformations a hierarchy of local, but staggered spin models with these criticalities were constructed. Here we show that this scheme can be simplified considerably by using another set of duality transformations that gives directly the translationally invariant $N$-cluster models considered in the main text.

The appropriately boundary coupled system of $N$ critical TFI chains is described by the Hamiltonian
\begin{align} \label{HN}
  H^{(N)}_{\rm cTFI}  = & \sum_{n=0}^{N-1} \left[ \sum_{j=0}^{L/N-2}  \left( \tau^x_{Nj+n} \tau^x_{N(j+1)+n} + \tau^z_{Nj+n} \right)  + \right. \nonumber \\
  \ & \qquad\left. \left( \prod_{m \neq n} \mathcal{P}_m \right) \tau^x_{L-N+n} \tau^x_{n}  + \tau^z_{L-N+n}
 \right],
\end{align}
where $\mathcal{P}_m = \prod_{j=0}^{L/N-1} \tau^z_{Nj+m}$ are the Ising symmetry operators of chain $m$. 
That this Hamiltonian is equivalent to the $N$-cluster model with $J=1$ and $h=0$ can be shown by using the following duality transformation 
\begin{align}
 \tau^z_{Nj} = & \sigma^y_{Nj} \left( \prod_{i=Nj+1}^{N(j+1)-2}  \sigma^z_i \right) \sigma^y_{N(j+1)-1}, \\
 \tau^z_{Nj+n} = & \sigma^x_{Nj-1+n}\sigma^x_{Nj+n},
\end{align}
and
\begin{align}
 \tau^x_{Nj} = & -\left( \prod_{m \neq 0} \mathcal{P}_m^{<j} \right) \sigma^x_{Nj}, \\
 \tau^x_{Nj+n} = & \left( \prod_{m < n} \mathcal{P}_m \right) \left( \prod_{m \neq n} \mathcal{P}_m^{<j} \right)  \left( \prod_{i=Nj}^{Nj-2+n} \sigma^z_{i} \right) \sigma^y_{Nj-1+n},
\end{align}
where now $n=1,\ldots,N-1$ and we have defined the string operators $\mathcal{P}_m^{<j} = \prod_{i<j} \tau^z_{Ni+m}$. Note that these $\mathcal{P}_m^{<j}$ are defined in terms of $\tau$ Pauli operators!
Inverting these duality transformations gives the transformations given in the main text for the $3$-cluster model.

On the other hand, the $N$-cluster model with $h=1$ and $J=0$ is obtained from $H^{(N-1)}_{\rm cTFI}$
by directly identifying $\sigma^z_{(N-1)j+n} = \tau^z_{(N-1)j+n}$ and introducing the dual operators
\begin{align}
	\tau^y_{(N-1)j+n} = & \left( \prod_{m < n} \mathcal{P}_m \right) \left( \prod_{m \neq n} \mathcal{P}_m^{< j} \right) \nonumber \\
	\ & \left( \prod_{i=Nj}^{Nj+n-1} \sigma^z_i \right) \sigma^y_{(N-1)j+n},
\end{align}
where now $\mathcal{P}_m^{<j} = \prod_{i<j} \tau^z_{(N-1)i+m}$, $j=0,\ldots,L/(N-1)-1$ and $n=0,\ldots,N-2$.

\subsection{Example: 3-cluster model with Ising perturbation}

To illustrate the general duality transformations, for the 3-cluster model with Ising perturbations they are explicitly given by
\begin{align}
 \tau^z_{3j} = & \sigma^y_{Nj} \sigma^z_{Nj+1} \sigma^y_{Nj+2}, \nonumber \\
 \tau^z_{3j+1} = & \sigma^x_{Nj}\sigma^x_{Nj+1}, \\
 \tau^z_{3j+2} = & \sigma^x_{Nj+1}\sigma^x_{Nj+2} \nonumber
\end{align}
and
\begin{align}
 \tau^x_{3j} = & \left( \prod_{i<j} \sigma^y_{Ni} \sigma^y_{Ni+1} \sigma^z_{Ni+2} \right) \sigma^y_{Nj} \sigma^y_{Nj+1} \sigma^y_{Nj+2}\nonumber \\ \ & \left( \prod_{i>j} \sigma^z_{Ni} \sigma^y_{Ni+1} \sigma^y_{Ni+2} \right), \nonumber \\
 \tau^x_{3j+1} = & \left( \prod_{i<j} \sigma^y_{Ni} \sigma^y_{Ni+1} \sigma^z_{Ni+2} \right) \sigma^y_{Nj}, \\
 \tau^x_{3j+2} = & \sigma^y_{Nj+2} \left( \prod_{i>j} \sigma^z_{Ni} \sigma^y_{Ni+1} \sigma^y_{Ni+2} \right) \nonumber
\end{align}
for $j=0,\ldots,L/3-1$. Inserting these into \rf{HN} with $N=3$ we obtain the translationally invariant critical Hamiltonian
\be
	H^{(3)}_{cluster} = \sum_{i=0}^{L-1} \sigma^y_{i} \sigma^z_{i+1} \sigma^y_{i+2} + \sigma^x_{i} \sigma^x_{i+1}
\ee
with a hidden $Z_2^{\times 3}$ global symmetry described by the symmetry operators
\begin{align}
 \mathcal{P}_0 = & \prod_{i=0}^{L/3-1} \sigma^y_{Ni} \sigma^z_{Ni+1} \sigma^y_{Ni+2}, \nonumber \\
 \mathcal{P}_1 = & \prod_{i=0}^{L/3-1} \sigma^x_{Nj}\sigma^x_{Nj+1}, \\
 \mathcal{P}_2 = & \prod_{i=0}^{L/3-1} \sigma^x_{Nj+1}\sigma^x_{Nj+2} \nonumber
\end{align}
that satisfy $\mathcal{P}_0\mathcal{P}_1\mathcal{P}_2=\prod_{i=0}^{L-1} \sigma^z_i$.

\section{The spectra of $so(N)_1$ critical points}

Conformal field theory gives a detailed prediction for the spectra of one-dimensional critical
systems \cite{byb}. When the finite-size spectrum is rescaled as described in the main text, the CFT predicts the energy levels and their degeneracies. This information is encoded in the
partition function of the CFT, which is composed of the partition functions of the left and right moving pieces,
one for each primary field. In general, for a primary field $\phi_i$, with scaling dimension $h_i$,
the left moving part of the partition function reads
\be
	Z_{l} (\phi_i) = q_{l}^{h_i} \sum_{n_l=0}^{\infty} c_{n_l} q^{n_l}_{l}
\ee
(and similar for the right moving part), where the $c_{n_l}$ are constants, depending on the primary field, and we
view $q_l$ as a formal variable. The total partition function
takes the form 
\be
	Z_{\rm tot} = \sum_{i} Z_{l} (\phi_{i}) Z_{r} (\phi_{i}),
\ee
where the sum runs over all primary fields in the theory. In the sector corresponding to primary field $\phi_i$, one can expand in powers of $q$ to obtain
\begin{equation}
\begin{split}
& Z_{l} (\phi_i) Z_{r} (\phi_i) = q_l^{h_i}q_r^{h_i} \cdot \\ &  (1 + c_{1,0} q_l + c_{0,1} q_r  + c_{2,0} q^2_l
+ c_{1,1} q_l q_r + c_{0,2} q^2_r + \cdots ).
\end{split}
\end{equation}
We assumed that the left and right scaling dimensions are equal: $h_{l,i} = h_{r,i} = h_{i}$, as is the
case for the CFTs we consider.
The coefficients $c_{n_l,n_r}=c_{n_l} c_{n_r}$ and the powers $n$ of the $q$ variables encode the spectrum as follows: Each term in this expansion correspond to $c_{n_l,n_r}$ degenerate states of energy $E=2h_i + n$.

The primary field content of $so(N)_1$ CFTs depends on the parity of $N$. For odd $N$ the $so(N)_1$ CFTs contain three primary fields $1$, $\psi$ and $\sigma$ with scaling dimensions $h_1=0$, $h_\psi=1/2$ and $h_\sigma=N/16$, respectively. On the other hand, for even $N$ there are four primary fields $1$, $\psi$, $\lambda$ and $\bar{\lambda}$ with scaling dimensions $h_1=0$, $h_\psi=1/2$ and $h_\lambda=h_{\bar{\lambda}}=N/16$, respectively. A thorough derivation of the partition functions of $so(N)_1$ CFTs can be found in Ref. \cite{Lahtinen14}. Here we just present the predictions for the few lowest lying energy levels and their degeneracies that are summarized in Tables \ref{Nodd}  and \ref{Neven}. For comparison, we also present in Table \ref{NIsing} the CFT prediction for the low lying energy levels of the ${\rm Ising}^{\times N}$ CFTs.

\begin{table}
\begin{tabular}{ccc}
N=1 & N=3 & N=5 \\
\begin{tabular}{|c|c|c|}
\hline
$\phi_i$ & $E$ & $c_{n_l,n_r}$ \\
\hline
\ & 0 & 1 \\
1 & 1 & 0 \\
\ & 2 & 2 \\
\hline
\ & 1 & 1 \\
$\psi$ & 2 & 2 \\
\ & 3 & 3 \\
\hline
\ & 1/8 & 1 \\
$\sigma$ & 9/8 & 2 \\
\ & 17/8 & 3 \\
\hline
\end{tabular} &
\begin{tabular}{|c|c|c|}
\hline
$\phi_i$ & $E$ & $c_{n_l,n_r}$ \\
\hline
\ & 0 & 1 \\
1 & 1 & 6 \\
\ & 2 & 27 \\
\hline
\ & 1 & 9 \\
$\psi$ & 2 & 24 \\
\ & 3 & 88 \\
\hline
\ & 3/8 & 4 \\
$\sigma$ & 11/8 & 24 \\
\ & 19/8 & 84 \\
\hline
\end{tabular} &
\begin{tabular}{|c|c|c|}
\hline
$\phi_i$ & $E$ & $c_{n_l,n_r}$ \\
\hline
\ & 0 & 1 \\
1 & 1 & 20 \\
\ & 2 & 160 \\
\hline
\ & 1 & 25 \\
$\psi$ & 2 & 150 \\
\ & 3 & 785 \\
\hline
\ & 5/8 & 16 \\
$\sigma$ & 13/8 & 160 \\
\ & 21/8 & 880 \\
\hline
\end{tabular}

\end{tabular}
\caption{The lowest lying energy levels $E=2h_i + n$ and their degeneracies $c_{n_l,n_r}$ for the primary field sectors $\phi_i \in \{ 1,\psi, \sigma \}$ of $so(N)_1$ CFTs with $N$ odd. The relevant scaling dimensions are $h_1=0$, $h_\psi=1/2$ and $h_\sigma=N/16$. Degeneracy of $c_{n_l,n_r}=0$ denotes no state at this energy. \label{Nodd} }
\end{table}

\begin{table}
\begin{tabular}{ccc}
N=2 & N=4 & N=6 \\
\begin{tabular}{|c|c|c|}
\hline
$\phi_i$ & $E$ & $c_{n_l,n_r}$ \\
\hline
\ & 0 & 1 \\
1 & 1 & 2 \\
\ & 2 & 9 \\
\hline
\ & 1 & 4 \\
$\psi$ & 2 & 8 \\
\ & 3 & 20 \\
\hline
\ & 1/4 & 1 \\
$\lambda, \bar{\lambda}$ & 5/4 & 4 \\
\ & 9/4 & 10 \\
\hline
\end{tabular} &
\begin{tabular}{|c|c|c|}
\hline
$\phi_i$ & $E$ & $c_{n_l,n_r}$ \\
\hline
\ & 0 & 1 \\
1 & 1 & 12 \\
\ & 2 & 70 \\
\hline
\ & 1 & 16 \\
$\psi$ & 2 & 64 \\
\ & 3 & 288 \\
\hline
\ & 1/2 & 4 \\
$\lambda, \bar{\lambda}$ & 3/2 & 32 \\
\ & 5/2 & 144 \\
\hline
\end{tabular}&
\begin{tabular}{|c|c|c|}
\hline
$\phi_i$ & $E$ & $c_{n_l,n_r}$ \\
\hline
\ & 0 & 1 \\
1 & 1 & 30 \\
\ & 2 & 327 \\
\hline
\ & 1 & 36 \\
$\psi$ & 2 & 312 \\
\ & 3 & 1900 \\
\hline
\ & 3/4 & 16 \\
$\lambda, \bar{\lambda}$ & 7/4 & 192 \\
\ & 11/4 & 1248 \\
\hline
\end{tabular}
\end{tabular}
\caption{The lowest lying energy levels $E=2h_i + n$ and their degeneracies $c_{n_l,n_r}$ for the sectors $\phi_i \in \{ 1,\psi,\lambda, \bar{\lambda} \}$ of $so(N)_1$ CFTs with $N$ even. The relevant scaling dimensions are $h_1=0$, $h_\psi=1/2$ and $h_\lambda=h_{\bar{\lambda}}=N/16$. The fields $\lambda$ and $\bar{\lambda}$ are spectrally indistinguishable, which means that the energy levels for these primary fields are degenerate by an additional factor of 2. \label{Neven}}
\end{table}

\begin{table*}[hb]
\begin{tabular}{cccccc}
N=1 & N=2 & N=3 & N=4 & N=5 & N=6\\
\begin{tabular}[t]{|c|c|}
\hline
$E$ & deg.\\
\hline
0 & 1\\
1 & 1\\
2 & 4\\
\hline
1/8 & 1\\
9/8 & 2\\
17/8 & 3\\
\hline
\end{tabular}
&
\begin{tabular}[t]{|c|c|}
\hline
$E$ & deg.\\
\hline
0 & 1\\
1 & 2\\
2 & 9\\
\hline
1/8 & 2\\
9/8 & 6\\
17/8 & 18\\
\hline
1/4 & 1\\
5/4 & 4\\
9/4 & 10\\
\hline
\end{tabular}
&
\begin{tabular}[t]{|c|c|}
\hline
$E$ & deg.\\
\hline
0 & 1\\
1 & 3\\
2 & 15\\
\hline
1/8 & 3\\
9/8 & 12\\
17/8 & 48\\
\hline
1/4 & 3\\
5/4 & 15\\
9/4 & 54\\
\hline
3/8 & 1\\
11/8 & 6\\
19/8 & 21\\
\hline
\end{tabular}
&
\begin{tabular}[t]{|c|c|}
\hline
$E$ & deg.\\
\hline
0 & 1\\
1 & 4\\
2 & 22\\
\hline
1/8 & 4\\
9/8 & 20\\
17/8 & 96\\
\hline
1/4 & 6\\
5/4 & 36\\
9/4 & 162\\
\hline
3/8 & 4\\
11/8 & 28\\
19/8 & 124\\
\hline
1/2 & 1\\
3/2 & 8\\
5/2 & 36\\
\hline
\end{tabular}
&
\begin{tabular}[t]{|c|c|}
\hline
$E$ & deg.\\
\hline
0 & 1\\
1 & 5\\
2 & 30\\
\hline
1/8 & 5\\
9/8 & 30\\
17/8 & 165\\
\hline
1/4 & 10\\
5/4 & 70\\
9/4 & 370\\
\hline
3/8 & 10\\
11/8 & 80\\
19/8 & 420\\
\hline
1/2 & 5\\
3/2 & 45\\
5/2 & 240\\
\hline
5/8 & 1\\
13/8 & 10\\
21/8 & 55\\
\hline
\end{tabular}
&
\begin{tabular}[t]{|c|c|}
\hline
$E$ & deg.\\
\hline
0 & 1\\
1 & 6\\
2 & 39\\
\hline
1/8 & 6\\
9/8 & 42\\
17/8 & 258\\
\hline
1/4 & 15\\
5/4 & 210\\
9/4 & 720\\
\hline
3/8 & 20\\
11/8 & 180\\
19/8 & 1080\\
\hline
1/2 & 15\\
3/2 & 150\\
5/2 & 915\\
\hline
5/8 & 6\\
13/8 & 66\\
21/8 & 414\\
\hline
3/4 & 1\\
7/4 & 12\\
11/4 & 78\\
\hline
\end{tabular}
\end{tabular}
\caption{The lowest lying energy levels $E<3$ and their degeneracies of the
${\rm Ising}^{\times N}$ CFTs, for $N=1,\ldots,6$. For each ${\rm Ising}^{\times N}$ CFT the spectrum splits into $3^N$ primary field sectors, which, for the sake of clarity, are not presented.} \label{NIsing}
\end{table*}

\end{document}